\documentclass[3p, preprint, twocolumn]{elsarticle}
\makeatletter
\def\ps@pprintTitle{%
\let\@oddhead\@empty
\let\@evenhead\@empty
\def\@oddfoot{\footnotesize\itshape
 \ifx\@journal\@empty Accepted for Publication in Elsevier Pattern Recognition Letters
\else\@journal\fi\hfill}%
\let\@evenfoot\@oddfoot}
\makeatother

\usepackage[hidelinks]{hyperref}
\usepackage{amsmath,amssymb,amsfonts}
\usepackage{booktabs}
\usepackage{multirow}
\usepackage{pifont}
\usepackage{url}
\usepackage{float}
\usepackage{caption}
\usepackage{subfig}
\usepackage{balance}
\biboptions{compress}

\def\x{{\mathbf x}}														% Definition of the bold letter x
\def\q{{\mathbf q}}														% Definition of the bold letter q
\def\p{{\mathbf p}}														% Definition of the bold letter p
\def\W{{\mathbf W}}														% Definition of the bold letter W
\def\ii{{\hat{\imath}}}												% Definition of the immaginary symbol i
\def\ij{{\hat{\jmath}}}												% Definition of the immaginary symbol j
\def\ik{{\hat{\kappa}}}												% Definition of the immaginary symbol k
\def\bH{\mathbb{H}}														% Definition of the mathbb letter H
\def\bR{\mathbb{R}}														% Definition of the mathbb letter R
\newcommand{\parallelsum}{\mathbin{\!/\mkern-5mu/\!}}  %% simbolo per reti parallele

\begin{document}

\begin{frontmatter}

\title{Dual Quaternion Ambisonics Array for \\Six-Degree-of-Freedom Acoustic Representation}

\author{Eleonora Grassucci$^{**}$, Gioia Mancini, Christian Brignone, Aurelio Uncini, and Danilo Comminiello}
\address{Dept. of Information Engineering, Electronics and Telecommunications (DIET), Sapienza University of Rome, Italy.}
\cortext[mycorrespondingauthor]{Corresponding author's email: eleonora.grassucci@uniroma1.it}

\begin{abstract}
Spatial audio methods are gaining a growing interest due to the spread of immersive audio experiences and applications, such as virtual and augmented reality. For these purposes, 3D audio signals are often acquired through arrays of Ambisonics microphones, each comprising four capsules that decompose the sound field in spherical harmonics. In this paper, we propose a dual quaternion representation of the spatial sound field acquired through an array of two First Order Ambisonics (FOA) microphones. The audio signals are encapsulated in a dual quaternion that leverages quaternion algebra properties to exploit correlations among them. This augmented representation with 6 degrees of freedom (6DOF) involves a more accurate coverage of the sound field, resulting in a more precise sound localization and a more immersive audio experience. We evaluate our approach on a sound event localization and detection (SELD) benchmark. We show that our dual quaternion SELD model with temporal convolution blocks (DualQSELD-TCN) achieves better results with respect to real and quaternion-valued baselines thanks to our augmented representation of the sound field. Full code is available at: \url{https://github.com/ispamm/DualQSELD-TCN}.
\end{abstract}

\begin{keyword}
Dual Quaternions, Quaternion Neural Networks, Quaternion Ambisonics Signals, Dual Quaternion Neural Networks
\end{keyword}

\end{frontmatter}

%\linenumbers

\section{Introduction}
\label{sec:intro}

In recent years, spatial audio is knowing an increasing attention also due to the widespread developing of applications requiring an immersive audio experience, such as virtual reality, scene characterization, speech enhancement or separation, and sound source localization \cite{Michelsanti2021TASLP, Manamperi2022TASLP, Comanducci2020TASLP, Cisneros2019VR}. Indeed, while virtual reality (VR) builds spaces different from real-life, augmented reality (AR) expands them through an enlarged user listening experience that is often based on the user immersion into a 3D sound field \cite{Kailas2021Springer, Majumder_2021_ICCV, Sundareswaran2003AR3D}. Frequently, this spatial sound field is acquired through First Order Ambisonics (FOA) 
microphones \cite{Saladnet2021, Mroz20216dof, Gotz2021Amb, guizzo2021l3das21, Guizzo2022L3DAS22CL}, which are arrays of four ideally coincident capsules that decompose the sound field into a combination of spherical harmonics. Due to the strong correlation among the four Ambisonics signals, quaternion neural networks (QNNs) have demonstrated their ability to suitable model these inputs and grasp information coming from each capsule. Indeed, QNNs handle the four signals as a single component and thanks to the quaternion algebra properties, in particular the Hamilton product, they preserve signals relations and correlations. Hypercomplex and quaternion models have shown interestingly results in different tasks such as 3D human motion prediction \cite{CAO2022141}, facial expression recognition \cite{Zhou2022TETCI}, image restoration/inpainting \cite{Huang2021QRestore, Jia2022TIP}, including sound event localization and detection (SELD), which is the task of jointly learning the temporal and spatial location of a sound and its class \cite{SELD, ComminielloICASSP2019b, Brignone2022ISCAS, QSSL, SalvatiIJCNN2020}, among others \cite{math10071083, NavarroMorenoR21, Chen2021Qfact, Guo2021QHyper, Zhang2022TNNLS}. Despite QNNs success in spatial source localization, no tests have been made with multiple Ambisonics microphones since these networks do not properly handle non-$4$D inputs due to the four-dimensional nature of quaternion numbers. However, an array of two Ambisonics provides a more precise sound field coverage, thus an appropriate processing of these features may bring better localization predictions \cite{Poschadel2021EUSIPCO}. Moreover, while the sound event detection (SED) sub-task is relatively easy, the sound direction of arrival (DOA) estimation sub-task strongly depends on the accuracy of the sound field reconstruction.

Recently, dual quaternion neural networks (DualQNNs) have been shown to be particularly suitable for modelling transformations in $3$D space. Due to the $6$ degrees of freedom (6DOF) of the the unit dual quaternion representation, DualQNNs can properly model rotations and translations of rigid bodies \cite{Poppelbaum2021DQRigid, Schwung2021RigidBody, Schilling2019HierarchicalDQ, TsiotrasDualQuat2020}, pose estimation/tracking \cite{Sveier2021DualQ, Gui2021DualQ}, and knowledge graph embeddings \cite{CaoDual2021, Nguyen2022NodeCB}. DualQNNs lay their foundations in dual quaternion numbers, which are a composition of two quaternions, thus particularly suitable for encapsulating eight-dimensional inputs.

For these reasons, we propose to exploit DualQNNs properties, including the ability to model 6DOF transformations in the $3$D space, to represent an array of two Ambisonics microphones, whose 6DOF \cite{Plinge2018SixDegreesofFreedomBA} perfectly fit with our augmented characterization. Thanks to the unit dual quaternion representation, our method is able to more precisely reconstruct and augment the spatial sound field, thus improving the localization capability of the model and the quality of user audio immersion in AR and VR applications. We evaluate the potential of our approach in the SELD task with a focus on the DOA sub-task in the recent benchmark Learning 3D Audio Sources (L3DAS21) dataset \cite{guizzo2021l3das21}. We show how the proposed 6DOF dual quaternion characterization for an array of two Ambisonics microphones better models the spatial sound field, thus achieving a more precise localization. Specifically, our contributions are:

\begin{itemize}
    \item We propose a dual quaternion representation for an array of Ambisonics microphones, which exploits quaternion algebra properties to preserve correlations among input signals and dual quaternion features to properly reconstruct the $3$D spatial sound field.
    \item We show how the proposed dual quaternion sound event localization and detection network (DualQSELD-TCN) presents an increased ability for the localization sub-task due to the dual quaternion representation.
    % \item We suggest a further approach to improve sound localization that jointly involves the dual quaternion representation and the features extracted by the generalized cross correlation function with phase transformation (GCC-PHAT). This method gains the best performance according to objective metrics in out tests.
    \item We build the global SELD (G-SELD) metrics, a more robust evaluation metrics for the SELD task, which balances the assessments computed with both the \textit{location-sensitive detection} metrics and the \textit{class-sensitive localization} metrics.  
\end{itemize}

The rest of the paper is organized as follows. In Section~\ref{sec:qnn} we expound the background on quaternion algebra and dual quaternion operations. Section~\ref{sec:qamb} introduces the dual quaternion Ambisonics representation, while Section~\ref{sec:method} presents the proposed DualQSELD-TCN, which is evaluated in Section~\ref{sec:exp}. Finally, we draw conclusions in Section~\ref{sec:con}.

\section{Quaternions and Dual Quaternions Background}
\label{sec:qnn}

\subsection{Quaternion numbers}

A quaternion number is a direct non-commutative extension of a complex-valued number, involving three imaginary units and four real-valued coefficients. More generally, the set of quaternion numbers $\bH$ lies in a four-dimensional associative normed division algebra, belonging to the class of Clifford algebras \cite{Ward1997}. A quaternion number is represented as

\begin{equation}
\label{eq:quaternion}
    q = q_W + q_X \ii + q_Y \ij + q_Z \ik,
\end{equation}

\noindent whereby $q_W$ is the real part and $\q = q_X \ii + q_Y \ij + q_Z \ik$ the imaginary one, in which the units comply with $\ii^2 = \ij^2 = \ik^2 = -1$, yielding to the multiplication $\ii \ij = - \ij \ii; \; \ii \ik = - \ik \ii; \; \ij \ik = - \ik \ij$. If the real part $q_W$ is equal to $0$, the resulting element is called a \textit{pure quaternion}. The conjugate of a quaternion is $q^* = q_W - q_X \ii - q_Y \ij - q_Z \ik$, while a quaternion with $\left\|q\right\|=1$, where $\left\|\cdot\right\|$ represents the Euclidean norm in $\bR^4$, is a unit quaternion. The addition of two quaternions $q$ and $p$ is performed element-wise as $q+p = (q_W + p_W) + (q_X + p_X) \ii + (q_Y + p_Y) \ij + (q_Z + p_Z) \ik$. The multiplication between two quaternions is instead defined by $q \otimes p = (q_W p_W - \q \cdot \p, \; \q \times \p + q_W \p + p_W \q)$, in which $\cdot$ is the dot product among vectors and $\times$ is the cross product. The multiplication between quaternions, also known as Hamilton product, was defined to properly model interplays among imaginary units due to the non-commutativity of cross products in this domain. Often, a quaternion is also implicitly indicated by 4D vectors involving its real representation that consider the four real-valued coefficients only. Therefore, leveraging this representation for the input quaternion $p = [p_W, p_X, p_Y, p_Z]$, the Hamilton product can be also expressed in a matrix-vector multiplication form as follows:

\begin{equation}
\label{eq:qprod}
    q \otimes p = \left[ {\begin{array}{*{20}c}
   \hfill {{q}_W } & \hfill { - {q}_X } & \hfill { - {q}_Y } & \hfill { - {q}_Z } \\
   \hfill {{q}_X } & \hfill {{q}_W } & \hfill { - {q}_Z } & \hfill {{q}_Y } \\
   \hfill {{q}_Y } & \hfill {{q}_Z } & \hfill {{q}_W } & \hfill { - {q}_X } \\
   \hfill {{q}_Z } & \hfill { - {q}_Y } & \hfill {{q}_X } & \hfill {{q}_W } \\
\end{array}} \right] \left[ {\begin{array}{*{20}c}
   {p_W } \hfill  \\
   {p_X } \hfill  \\
   {p_Y } \hfill  \\
   {p_Z } \hfill  \\
\end{array}} \right].
\end{equation}

Moreover, quaternions are particularly appropriate to represent rotation in $\bR^3$. Indeed, a unit quaternion can be expressed through polar coordinates as 

\begin{equation}
\label{eq:qpolar}
    q_{\theta} = q_0 + \mathbf{q} = \cos\theta + \mathbf{u}\sin\theta,
\end{equation}

\noindent with $\theta \in (-\pi, \pi]$ and $\mathbf{u}$ unit vector that indicates the direction. The rotation matrix can be then defined following \cite{Kuipers1921book}, and applied to the unit quaternion vector to be rotated.

\subsection{Dual quaternion numbers}

Dual numbers have a similar form to complex numbers and still being part of a hypercomplex number system discovered by Clifford \cite{Clifford1871}. They are composed of two elements, a real part and a dual part multiplied by the dual unit. Formally, they can be introduced as

\begin{equation}
\label{eq:dualnum}
    \hat{d} = d_1 + \epsilon d_2,
\end{equation}

\noindent with $d_1, d_2 \in \bR$ and $\epsilon$ dual unit complying with the properties $\epsilon \neq 0, \; \epsilon^2 = 0$. As for quaternions, also this domain involves the conjugate operation, which is described as $\hat{d}^* = d_1 - \epsilon d_2$. Addition and subtraction of dual numbers are element-wise operations: $\hat{d} + \hat{c} = (d_1 + c_1) + \epsilon (d_2 + c_2)$, while the multiplication is $\hat{d}\hat{c} = d_1c_1 + \epsilon (d_1c_2 + c_1d_2) + \epsilon^2 (c_2d_2)$, however, since $\epsilon^2=0$, the last term vanishes. As for the Hamilton product, also dual numbers product has a matrix form which follows

\begin{equation}
\label{eq:dualprod}
    \hat{d}\hat{c} = \left[ {\begin{array}{*{20}c}
    \hfill {c_1d_2} & \hfill \hfill {0} \\
   \hfill {d_1c_1} & {d_1c_2} \\
    \end{array}} \right].
\end{equation}

Dual quaternions are dual numbers involving quaternions instead of real coefficients with the dual unit $\epsilon$ that commutes with each element of the algebra that is $\epsilon \ii = \ii \epsilon; \; \epsilon \ij = \ij \epsilon; \; \epsilon \ik = \ik \epsilon$. Differently from quaternions, however, they do not form a division algebra. Given two quaternions $q, q_{\epsilon} \in \bH$, a dual quaternion number is expressed as

\begin{equation}
    \hat{q} = q + \epsilon q_{\epsilon}.
\end{equation}

Interestingly, due to their complexity, dual quaternions have two conjugation operations. The first one is computed by conjugating both the quaternions $q$ and $q_{\epsilon}$ resulting in $\hat{q}^{*_1} = q_W - q_X \ii - q_Y \ij - q_Z \ik + \epsilon(q_{\epsilon, W} - q_{\epsilon, X} \ii - q_{\epsilon, Y} \ij - q_{\epsilon, Z} \ik)$. The second conjugation is computed by conjugating the dual unit too, thus the formula becomes $\hat{q}^{*_2} = q_W - q_X \ii - q_Y \ij - q_Z \ik + \epsilon(-q_{\epsilon, W} + q_{\epsilon, X} \ii + q_{\epsilon, Y} \ij + q_{\epsilon, Z} \ik)$. As for previous number systems, also dual quaternions involve an element-wise addition: $\hat{q} + \hat{p} = (q+p) + \epsilon (q_{\epsilon} + p_{\epsilon})$. The multiplication operation can be instead introduced involving the Hamilton product in \eqref{eq:qprod} as $\hat{q} \hat{p} = q \otimes p + \epsilon (q\otimes p_{\epsilon} + q_{\epsilon} \otimes p)$. Therefore, its matrix form can be written as:

\begin{equation}
\label{eq:dualqprod}
    \hat{q} \hat{p} = \left[ {\begin{array}{*{20}c}
     \hfill {(q \otimes p_{\epsilon})} & \hfill {0} \\
     \hfill {q \otimes p} &  {(q_{\epsilon} \otimes p)} \\
    \end{array}} \right],
\end{equation}

\noindent whereby $\otimes$ is the Hamilton product in \eqref{eq:qprod} and the upper-right term is $0$ due to the dual unit property $\epsilon^2=0$.

Interestingly, dual quaternions are suitable for jointly applying rotation and translation. Given a quaternion in polar form $q_\theta$ as in Eq.\ref{eq:qpolar}, the corresponding rotation matrix $R$ and a translation vector $\mathbf{t}=(t_1, t_2, t_3)$, a point $v$ can be rotated and then translated by $Rv+t$ \cite{CaoDual2021, Jia2018DualQ}. Then, the rotation-translation transformation can be encapsulated in a dual quaternion:

\begin{equation}
    \hat{\sigma} = q_{\theta} + \frac{\epsilon}{2} \mathbf{t}q_{\theta}.
\end{equation}

\section{Dual Quaternion Ambisonics Signals}
\label{sec:qamb}

First-order Ambisonics (FOA) microphones are composed of 4 capsules in which the first one corresponds to the spherical harmonic of order $0$ and it is usually a pressure microphone that is an omnidirectional microphone. The latter three capsules capture instead the acoustic velocity, corresponding to the harmonic functions of order $1$. Respectively, these capsules are named $W, X, Y,$ and $Z$. Therefore, a discrete-time signal $s[n]$ with angles $\theta, \phi$ can be represented in the so-called B-format Ambisonics representation:

\begin{equation}
\label{eq:qamb}
\begin{cases}
     x_{W}[n] = s[n]/\sqrt{3} \\
     x_{X}[n] = s[n]\cos{\theta}\cos{\phi} \\
     x_{Y}[n] = s[n]\sin{\theta}\cos{\phi} \\
     x_{Z}[n] = s[n]\sin{\phi}.
\end{cases}
\end{equation}

This representation involves highly correlated components that can be straightforwardly enclosed in a quaternion by treating the four signals as the real-valued coefficients of the quaternion \cite{ComminielloICASSP2019b, QSSL}. However, when dealing with an array of two microphones, we need to encapsulate the resulting eight channels in two different quaternions, where each microphone signal will have the form in \eqref{eq:qamb}. Nevertheless, these representations are processed as different entities since QNNs deal with four-dimensional inputs only. This separation may cause a loss of correlated information coming from the eight capsules signal. An array of two Ambisonics microphones is usually employed to have a more wide and precise coverage of the spatial sound field. Nevertheless, using two different quaternions to treat them just build two different spatial representations, while the proper way should be building an augmented representation of it.

To address this issue, we propose to exploit the augmented representation given by the dual quaternion form and enclose the B-format Ambisonics signals of microphones $A$ and $B$ in
\begin{equation}
\label{eq:dqamb}
\begin{split}
    \hat{x} = \; &x_{W}^{A}[n] + x_{X}^{A}[n] \ii + x_{Y}^{A}[n] \ij + x_{Z}^{A}[n] \ik \\
    &+ \epsilon (x_{W}^{B}[n] + x_{X}^{B}[n] \ii + x_{Y}^{B}[n] \ij + x_{Z}^{B}[n] \ik).
\end{split}
\end{equation}

Thanks to the dual quaternion characterization, we build a more compact representation of the two microphones signals. However, \eqref{eq:dqamb} has eight degrees of freedom (8DOF) while the dual ambisonics spatial field has just six degrees. Therefore, to parameterize the spatial sound field, we have to impose some constraints. A common approach in kinematics, where known methods want to parameterize rotations (3DOF) and translations (3DOF), is normalizing \eqref{eq:dqamb} to a unit dual quaternion \cite{TsiotrasDualQuat2020, Schwung2021RigidBody, Poppelbaum2021DQRigid, Samanc2021TheNS}. Unit dual quaternions have unital norm as $\left\|\hat{x}\right\| = \hat{x} \cdot \hat{x} =1$. However, as introduced in Section~\ref{sec:qnn}, the product of two dual quaternions is composed of two terms, being $\hat{x} \cdot \hat{x} = x^A \cdot x^A + \epsilon(2 x^A \cdot x^B)$, without reporting the term multiplied by $\epsilon^2$ that vanishes due to the dual unit property. Therefore, to guarantee that such product is equal to $1$ we have to impose the two following constraints \cite{CaoDual2021, Valverde2018DynamicMA}.
\begin{align}
\label{eq:1const}
    x^A \cdot x^A &= x_{W}^{A^{2}} + x_{X}^{A^{2}} + x_{Y}^{A^{2}} + x_{Z}^{A^{2}} = 1, \\
    x^A \cdot x^B &= x_{W}^{A}x_{W}^{B} + x_{X}^{A}x_{X}^{B} + x_{Y}^{A}x_{Y}^{B} + x_{Z}^{A}x_{Z}^{B} = 0.
\end{align}

\noindent Therefore, we can satisfy the equations in \eqref{eq:1const} by exploiting the following formulations on the quaternions $x^A$ and $x^B$:

% Therefore, following these approaches \cite{CaoDual2021, Valverde2018DynamicMA, Poppelbaum2021DQRigid}, we denote

\begin{align}
    \bar{x}^{A} &= \frac{x^{A}}{\left\| x^{A}\right\|} = \frac{x_W^{A} + x_X^{A} + x_Y^{A} + x_Z^{A}}{\sqrt{x_{W}^{A^2} + x_{X}^{A^2} + x_{Y}^{A^2} + x_{Z}^{A^2}}}, \\
    \bar{x}^{B} &= x^{B} - \frac{x^{B}x^{A}}{\left\|x^{A}\right\|^2} x^{A}.
\end{align}

\noindent Then, the normalized form of the dual quaternion in \eqref{eq:dqamb} will involve the above-computed $\bar{x}^{A}$ and $\bar{x}^{B}$.
% From this formulation, we can deduce the two constraints to reduce the degrees of freedom \cite{CaoDual2021, Samanc2021TheNS}:
By applying these transformations to the dual quaternion input we reduce it to have 6DOF \cite{McCarthyDualQ1990}. This representation allows an accurate reconstruction of the spatial sound field leading to a more precise localization in the 3D space and a more immersive audio experience for AR and VR applications.

\section{Dual Quaternion Network for Sound Event Localization and Detection}
\label{sec:method}

\subsection{Quaternion and Dual Quaternion Networks}

In this section, we expound the main concepts underlying quaternion and dual quaternion neural networks. We give formal definitions for fully-connected (FC) layers in real, quaternion and dual quaternion domain, however, the same definitions hold for convolutional layers too \cite{GrassucciICASSP2021, GrassucciQGAN2021}.

Given an input $\mathbf{x} \in \bR^{m \times 1}$ and a set of parameters $\mathbf{W} \in \bR^{n\times m}$ of weights and $\mathbf{b} \in \bR^{n \times 1}$ biases, a real-valued FC layer takes the form

\begin{equation}
\label{eq:fclayer}
    \mathbf{y} = \sigma(\mathbf{W}\mathbf{x} + \mathbf{b}),
\end{equation}

\noindent whereby $\mathbf{y} \in \bR^{n \times 1}$ is the output, $\mathbf{W}$ contains $n \times m$ parameters, and $\sigma$ is the activation function.

Quaternion Neural Networks (QNNs) define each input, weight, bias and output as a quaternion in \eqref{eq:quaternion}. Therefore, the multiplication $\mathbf{W}\mathbf{x}$ in \eqref{eq:fclayer} becomes a multiplication between two quaternions and has to be performed following the Hamilton product in \eqref{eq:qprod}. Given a quaternion input $\mathbf{x} = \mathbf{x}_W + \mathbf{x}_X \ii + \mathbf{x}_Y \ij + \mathbf{x}_Z \ik$, the quaternion FC (Q-FC) layer with a weight matrix $\W = \W_W + \W_X \ii + \W_Y \ij + \W_Z \ik$, bias $\mathbf{b} = \mathbf{b}_W + \mathbf{b}_X \ii + \mathbf{b}_Y \ij + \mathbf{b}_Z \ik$ is then:

\begin{equation}
    \mathbf{y} = \sigma(\mathbf{W} \otimes \mathbf{x} + \mathbf{b}),
\end{equation}

\noindent in which the activation function $\sigma$ may be a quaternion function or a split activation function \cite{GrassucciQGAN2021}. The weight matrix $\W \in \bH^{n \times m}$ comprises real-valued submatrices, which are composed according to \eqref{eq:qprod}. Due to the reusing of submatrices, a quaternion weight matrix has $n \times m /4$ parameters, thus, quaternion layers save $75\%$ of parameters. Moreover, since submatrices are also shared among input components (e.g., $\W_W$ is multiplied for each one of the components of $\x$, and so on), QNNs are capable of learning correlations among input dimensions such as pixels of RGB images or signals of multichannel audio \cite{ComminielloICASSP2019b}.

As for quaternion models, also Dual Quaternion neural networks (DualQNNs) operate with dual quaternion inputs, weights and outputs. Therefore, the weight matrix $\W$ takes the form of \eqref{eq:dualqprod} involving two quaternion weight matrices $\mathbf{Q} = \mathbf{Q}_W + \mathbf{Q}_X \ii + \mathbf{Q}_Y \ij + \mathbf{Q}_Z \ik$ and $\mathbf{Q}_{\epsilon} = \mathbf{Q}_{\epsilon,W} + \mathbf{Q}_{\epsilon,X} \ii + \mathbf{Q}_{\epsilon,Y} \ij + \mathbf{Q}_{\epsilon,Z} \ik$. Interestingly, the resulting matrix is more sparse with respect to real and quaternion-valued matrices due to the zero-block component in \eqref{eq:dualqprod} \cite{Schwung2021RigidBody}. Due to this property, dual quaternion layers can be implemented to save computations, without multiplying the zero-part of the weight matrix, thus reducing the number products. However, this advantage strongly depends on the implementation since two different pathways can be covered. The first one relies on the building of the weight matrix including the zero-part as in \eqref{eq:dualqprod} and then multiply/convolve the full matrix for the input. The second alternative relies in splitting the multiplication in \eqref{eq:dualqprod}, thus avoiding the product between the zero-part and the input, thus saving computations.

\subsection{DualQSELD-TCN}

\begin{figure*}[t]
\centering
\subfloat[][\emph{DualQ-Conv-TC Block}]
{\includegraphics[trim= 0 0 0 0,clip,height=.35\textwidth]{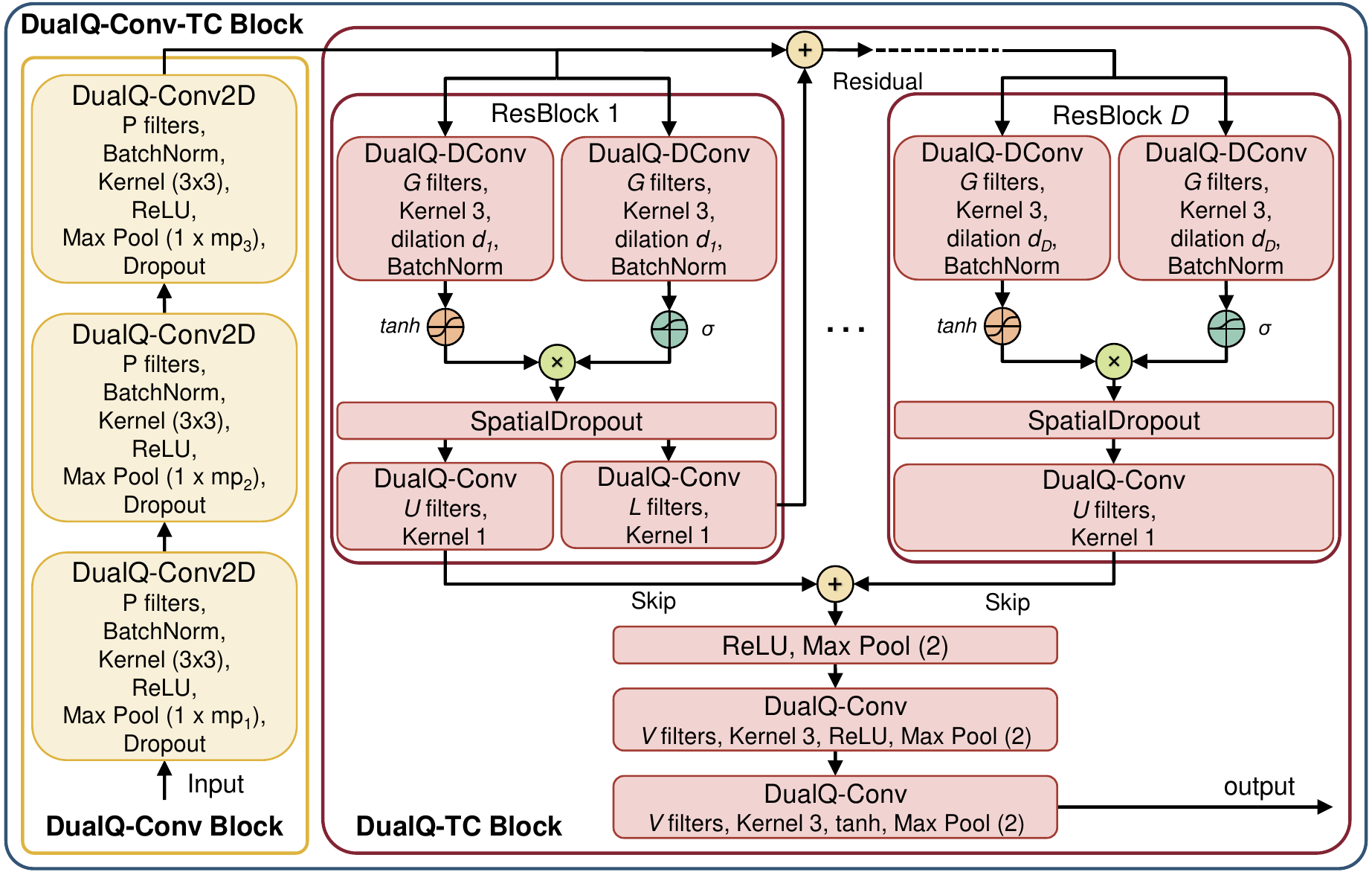}} 
\quad
\subfloat[][\emph{DualQSELD-TCN}]
{\includegraphics[trim= 0 0 0 0,clip,height=.35\textwidth]{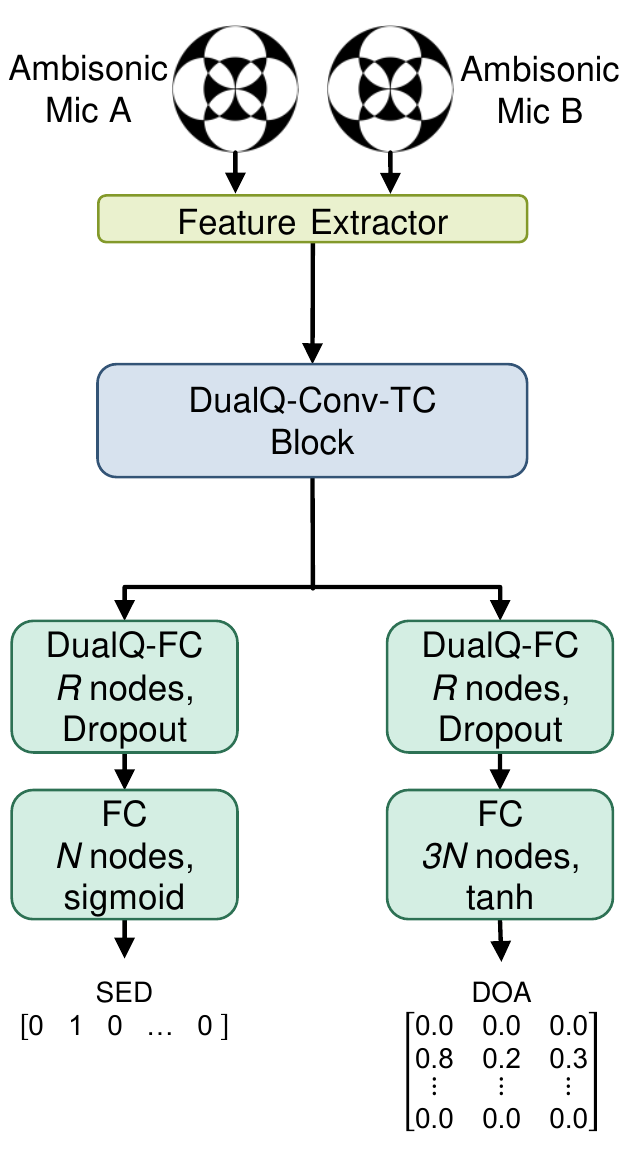}}
\quad
\subfloat[][\emph{DualQSELD-TCN$_{\parallelsum}$}]
{\includegraphics[trim= 0 0 0 0,clip,height=.35\textwidth]{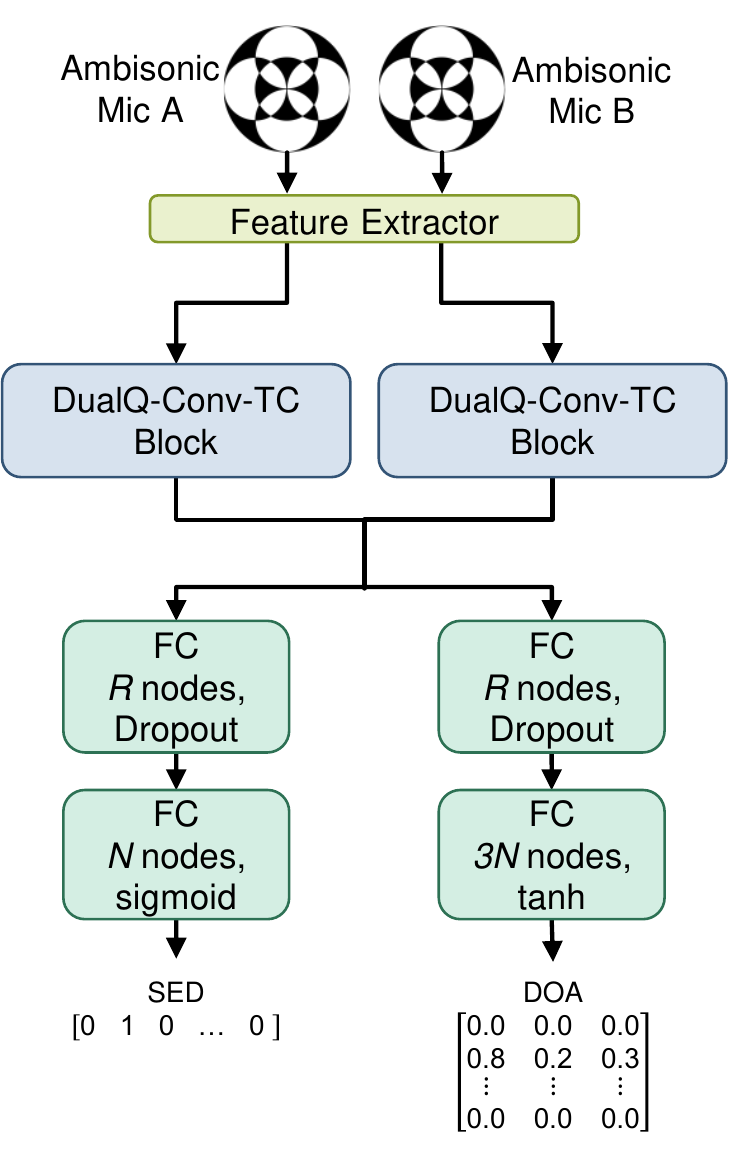}}
\caption{The proposed DualQSELD-TCN model. The DualQ-Conv-TC block (a) involves an initial DualQ convolution blocks and $10$ residual blocks with DualQ temporal convolutions and gated tanh unit. The overall architecture (b) takes the Ambisonics signals in input, extracts the features and pass them to the DualQ-Conv-TC block. Finally, two parallel branches give the predictions for the SED and DOA sub-task.}
\label{fig:arch}
\end{figure*}

To test the dual quaternion representation abilities in modelling a real-world 3D spatial sound field, we perform the task of sound event localization and detection (SELD). This task consists in jointly performing two sub-tasks: the first one is the sound event detection (SED) that aims at learning the class of the sound and the segment of the audio frame at which this sound appears, while the second one is the direction of arrival (DOA) estimation, thus learning the spatial coordinates of the sound source. The latter is the most arduous challenge and the one on which we focus our attention. Indeed, while the detection task is not heavily influenced by the kind of input representation we employ, a dual quaternion model should instead better capture the DOA thanks to its augmented representation of the $3$D space.

To this end, we propose a dual quaternion sound event localization and detection network with temporal convolutional blocks (DualQSELD-TCN) inspired by \cite{Brignone2022ISCAS}. In this model, we exploit operations in the dual quaternion domain, including the product in \eqref{eq:dualqprod} for neural layers and we enclose the two microphones signals as in \eqref{eq:dqamb}.
We extract the features through a short-time Fourier transform (STFT) with an Hamming window of length $512$. Magnitudes from the two microphones are then stacked resulting in an input of $T \times 256 \times 8$ with $T$ frames and where the last dimension becomes $16$ whether we involve also the phase information. The model, whose overall architecture is displayed in Fig.~\ref{fig:arch}b, then handles the eight-channel input through a dual quaternion convolution (DualQ-Conv) block and a DualQ temporal convolution (DualQ-TC) block, which together compose a DualQ-Conv-TC block (Fig.~\ref{fig:arch}a). Instead, the processing of the sixteen-channel input is performed by employing two parallel DualQ-Conv-TC blocks (Fig.~\ref{fig:arch}c), allowing to encapsulate the magnitude and the phase of each microphone in a dual quaternion and process them appropriately. Finally, in both cases, two separate branches take care of predicting the SED and the DOA. More in detail, the first architecture block comprises three DualQ-Conv2D layers, each followed by batch normalization layer, ReLU activation function, max pooling applied to the frequency axis and dropout. The input should then be properly organized to be fed into the DualQ-TC block, therefore we stack frequency and channel axis together \cite{Brignone2022ISCAS}. The core part of our model is the TC block, which has been introduced to remove the computational-heavy sequential processing of recurrent networks \cite{SELD-TCN}. It involves $10$ residual blocks containing DualQ non-causal dilated convolutions (DualQ-DConv) that enlarge the receptive field considering also future samples rather than only present and past ones \cite{oord2016wavenet, wavenet}. These layers have a dilation rate computed by employing the first ten numbers of the Fibonacci sequence. A batch normalization layer is applied to the output of each DualQ-DConv and, consequently, we employ a gated tanh unit (GTU) with the form
\begin{equation}
    \centering
    y=\tanh(\mathbf{W}_f * \mathbf{x}) \odot \sigma (\mathbf{W}_g * \mathbf{x}),
\end{equation}
\noindent in which $\mathbf{W}_f$ and $\mathbf{W}_g$ are the convolutional filters of the left and right side of the gate, with tanh and sigmoid activation functions, respectively. The input goes then into a spatial dropout and in two parallel 1D DualQ-Conv layers, where the first one is used for the skip connection and the second one to preserve the dimensionality of the residual connection that flows into the next residual block together with the original input. The sum of all the skip connections is then passed through a ReLU activation function and then to the last DualQ-Conv layers with ReLU and tanh, respectively, with max pooling applied after each activation in order to match the correct output dimension. Finally, in the eight-channel input case, the classifier branches are composed of DualQ fully connected layers (DualQ-FC) while, in the sixteen-channel input case, the classifier branches are completely real-valued. This is because the two parallel DualQ-TC blocks learn different representations and a dual quaternion representation of them would not be appropriate. In both cases, dropout is applied and an ending real-valued FC is employed to specify the number of classes for SED and the cartesian coordinates for DOA.

%Finally, the classifier branches are composed of DualQ fully connected layers (DualQ-FC), dropout and an ending real-valued FC to specify the number of classes for SED and the cartesian coordinates for DOA.

%\begin{figure*}
%    \centering
%    \includegraphics[width=\linewidth]{Figures/DQSELD-TCN__full.pdf}
%    \caption{Dual Quaternion SELD-TCN (DQSELD-TCN) model.}
%    \label{fig:arch}
%\end{figure*}

\section{Experiments and Discussion}
\label{sec:exp}

In this section, we present the experimental evaluation and the results discussion. To be consistent with previous literature where novel hypercomplex models are usually compared with their real-valued and quaternion-valued counterparts \cite{grassucci2021lightweight}, we evaluate the performance of the proposed approach against the real-valued SELD-TCN baseline \cite{SELD-TCN} and its quaternion counterpart, QSELD-TCN \cite{Brignone2022ISCAS}. We slightly modify the latter to properly process signals from two Ambisonics, thus we create two parallel Q-Conv-TC blocks in order to involve a quaternion for each microphone, resulting in an architecture similar to the DualQSELD-TCN parallel in Fig. \ref{fig:arch}c but using quaternion-valued layers instead of dual quaternion-valued layers. Moreover, in order to guarantee a fair comparison among models, we also test an augmented version of the QSELD-TCN with a number of parameters comparable to both the real-valued baseline and the proposed DualQSELD-TCN.

\subsection{L3DAS21 Dataset}
The Learning $3$D Audio Sources 2021 (L3DAS21) dataset for the SELD task contains approximately 15 hours of simulated office environment sounds divided in 900 1-minute-long samples, obtained through two MSMP B-format Ambisonics \cite{guizzo2021l3das21}. In each frame, there may be up to three simultaneously active sounds belonging to the $14$ sound classes selected from the FSD50K dataset \cite{Fonseca2022FSD50KAO}. Samples have a frequency of $32$ kHz, $16$ bit and the difference of amplitude among sounds ranges from $0$ to $20$ dBFS. The target consists of a matrix with dimension $[n\_frame, n\_class \times n\_overlap]$ for the SED sub-task, while for DOA sub-task we have a larger matrix of $[n\_frame, n\_class \times n\_overlap \times 3]$, where $3$ is for the three $x,y,z$ coordinates in the $3$D space, $n\_overlap=3$ and $n\_class=14$.

\begin{table*}[]
\caption{Metrics results on the L3DAS21 dataset. The first section reports the comparisons with real-valued and quaternion-valued baselines, while the last section involves further experiments with the proposed approach to improve the spatial sound field representation.}
\label{tab:metrics2}
\centering
\begin{tabular}{l|cc|ccc}
\toprule
Model & Params & Features & LSD$_{score}\downarrow$ & CSL$_{score}\downarrow$ & G-SELD$_{score}\downarrow$ \\\hline 
%QSELD-TCN & Mag & 0.523 & 0.555 & 84.62 & 0.634 & 0.467 & 0.516 & 407.464\\ 
SELD-TCN & 1.6M & Mag & 0.533 & 0.413 & 0.473 \\ %1.547.560
QSELD-TCN & 0.8M & Mag & \textbf{0.506} & 0.404 & 0.455 \\ %841.000
QSELD-TCN & 1.6M & Mag & 0.550 & 0.378 & 0.464 \\ %841.000
% DualQSELD-TCN old & Mag & 0.464 & 0.638 & 85.65 & 0.534 &0.529 & 0.587 & 217.448 \\
%DualQSELD-TCN(R) & Mag  &  0.516 & 0.412& 0.464 & 841.000\\
%DualQSELD-TCN & Mag &  &  &  & \\
DualQSELD-TCN & 1.8M & Mag & 0.512 & \textbf{0.365} &  \textbf{0.439} \\ %1.788.648
%DualQSELD-TCN L & Mag & \ding{43} &  &  &  &  &  &  &  \\
\hline
%SELD-TCN & Mag+Phase & 0.647 & 0.393 & 40.157 & 0.674 & 0.323 & 0.373 & 0.275 & 1.552.168 \\
%QSELD-TCN$_{\parallelsum}$ & Mag+Phase & 0.603 & 0.447 & 51.376 & 0.664 & 0.366 & 0.422 & 0.311 & \\
%DualQSELD-TCN & Mag+Phase &  &  &  &  &  &  &  & \\
DualQSELD-TCN & 1.8M & Mag+Phase &  0.410 & 0.303 & 0.356 \\ %1.790.376
DualQSELD-TCN$_{\parallelsum}$ & 3.6M & Mag+Phase &  \textbf{0.369} & \textbf{0.279} & \textbf{0.324} \\ %3.591.208
\bottomrule
\end{tabular}
\end{table*}

\subsection{Metrics}
To evaluate the performance of our model in a robust way, we employ several objective metrics. We compute the F score as suggested by the L3DAS21 Challenge and the error rate (ER) through the \textit{location-sensitive detection metrics} \cite{guizzo2021l3das21, Guizzo2022L3DAS22CL}. We build a score from these metrics, which we name location-sensitive detection (LSD$_{score}$), defined as
\begin{equation}
    \text{LSD}_{score} = \frac{\text{ER}+\left(1-\text{F}\right)}{2}.
\end{equation}
Then, we consider other two metrics suggested by the DCASE21 Challenge \cite{Politis2021OverviewAE}, the localization error (LE) and the localization recall (LR), which are computed through the \textit{class-sensitive localization metrics} \cite{Mesaros2019Measurement}. We give rise to another score from the latter ones, named class-sensitive localization (CSL$_{score}$), of the form
\begin{equation}
    \text{CSL}_{score} = \frac{\left(\text{LE}/180\right)+\left(1-\text{LR}\right)}{2}.
\end{equation}
Finally, we also involve a novel global metrics that combines LSD$_{score}$ and CSL$_{score}$, thus taking into account both the location-sensitive detection and the class-sensitive localization and then being more robust to unbalanced results. Indeed, the SELD task comprises two sub-tasks SED and DOA and an unsuitable or insufficient assessment of the two may compromise the global evaluation of the model. We name the average of LSD$_{score}$ and CSL$_{score}$ as global SELD (G-SELD$_{score}$). The closer to $0$ the value of the G-SELD$_{score}$, the better the predictions are, while values equal to $1$ indicates bad performance. We believe that the G-SELD$_{score}$ is a more robust evaluation metrics, being based on four different metrics that are computed with two diverse evaluation methods.

\subsection{Architecture and training}
In this subsection, we provide training and architectures details to reproduce our experiments.

We test two versions of the proposed DualQSELD-TCN network, named DualQSELD-TCN (Fig.~\ref{fig:arch}b) and DualQSELD-TCN parallel (DualQSELD-TCN$_{\parallelsum}$) in Fig.~\ref{fig:arch}c, the latter with two parallel DualQ-Conv-TC blocks in order to encapsulate the magnitudes and phases features of each microphone in two different dual quaternions.
The proposed model comprises the initial convolutional blocks with $P = 192$ filters of dimension $3 \times 3$ for each layer and pooling of $mp = [8, 8, 2]$, respectively, while the dropout probability is $0.3$. In the DualQ-TC block, where we stack $D = 10$ resblocks resulting in a receptive field of $287$, the DualQ dilated convolutions involves $G = 384$ filters with size $3$, while both the parallel 1D DualQ-Conv have a kernel size equal to $1$ and $U = L = 384$ and the spatial dropout probability is set to $0.5$. The last two DualQ-Conv of the DualQ-TC block still have $V = 384$ filters of size $3$. Finally, the classifier branches are then composed of DualQ-FC layers of $R = 384$ nodes each (FC layers of $R = 128$ nodes each in the case of DualQSELD-TCN$_{\parallelsum}$), with the final real-valued layer with a number of nodes equal to the required target. We train each model for a minimum of $1000$ epochs and an early stopping with patience equal to $300$, through the Adam optimizer having an initial learning rate of $0.0001$. We jointly optimize the SED binary cross-entropy loss and the DOA mean squared error by weighting the latter $5$ times with respect to the first one \cite{guizzo2021l3das21}.

\subsection{Results discussion}
We perform the tests in two scenarios. First, we involve as input to the model the magnitudes extracted from the two Ambisonics microphones, thus the eight channels are encapsulated in the dual quaternion as in \eqref{eq:dqamb}. Then, we want to further push our approach leveraging also the phase information, thus passing to the model a sixteen-channel input.

In the first section of Table~\ref{tab:metrics2}, we report the computed scores for the comparison with the baselines SELD-TCN and QSELD-TCN in the real and quaternion domain, respectively. The quaternion network performs better than the real-valued one, indicating the advantages of processing Ambisonics signals through quaternion algebra, as stated in \cite{Brignone2022ISCAS, ComminielloICASSP2019b, QSSL}. Our approach, exceeds both the baselines, especially in the CSL$_{score}$, thus having better LE and LR scores. This means that our DualQSELD-TCN model improves the sound localization with respect to the baselines, thanks to the augmented representation of the dual quaternion. Once we ensure that the proposed model can build a more accurate representation of the sound field, we further push up its capabilities considering also the phase information as input features. This results in more accurate scores for both the tasks with a G-SELD$_{score}$ improved from $0.439$ to $0.356$. Finally, we achieve the best scores with the parallel version DualQSELD-TCN$_{\parallelsum}$, which encapsulates magnitudes and phases of each microphone in two diverse dual quaternion representations. This enhanced model gains the best G-SELD$_{score}$ of $0.324$.

\section{Conclusion}
\label{sec:con}

In this paper, we introduce a novel augmented representation for the spatial sound field acquired through an array of Ambisonics microphones. Our dual quaternion representation exploits the quaternion algebra to preserve correlations among signals, while building an augmented characterization of the sound field with six degrees of freedom. We show the improved abilities of our approach in the sound event localization and detection (SELD) task, where the proposed dual quaternion SELD-TCN (DualQSELD-TCN) network outperforms both real and quaternion-valued baselines.

\section*{Acknowledgements}

This research was funded by ``Progetti di Ricerca'' of Sapienza University of Rome under grant numbers RM120172AC5A564C and RG11916B88E1942F. We would also like to thank the NVIDIA Applied Research Accelerator Program for the donation of a NVIDIA Quadro RTX 8000 for the project ``Quaternion Deep Learning for 3D Audio Sources''.

\balance
%\bibliography{DQnetbiblio}

\begin{thebibliography}{10}
\expandafter\ifx\csname url\endcsname\relax
  \def\url#1{\texttt{#1}}\fi
\expandafter\ifx\csname urlprefix\endcsname\relax\def\urlprefix{URL }\fi
\expandafter\ifx\csname href\endcsname\relax
  \def\href#1#2{#2} \def\path#1{#1}\fi

\bibitem{Michelsanti2021TASLP}
D.~Michelsanti, Z.~H. Tan, S.~X. Zhang, Y.~Xu, M.~Yu, D.~Yu, J.~Jensen, An
  overview of deep-learning-based audio-visual speech enhancement and
  separation, {IEEE/ACM} Trans. on Audio, Speech, and Language Process. 29
  (2021) 1368--1396.

\bibitem{Manamperi2022TASLP}
W.~Manamperi, T.~D. Abhayapala, J.~Zhang, P.~N. Samarasinghe, Drone audition:
  Sound source localization using on-board microphones, {IEEE/ACM} Trans. on
  Audio, Speech, and Language Process. 30 (2022) 508--519.

\bibitem{Comanducci2020TASLP}
L.~Comanducci, F.~Borra, P.~Bestagini, F.~Antonacci, S.~Tubaro, A.~Sarti,
  Source localization using distributed microphones in reverberant environments
  based on deep learning and ray space transform, {IEEE/ACM} Trans. on Audio,
  Speech, and Language Process. 28 (2020) 2238--2251.

\bibitem{Cisneros2019VR}
R.~Cisneros, K.~Wood, S.~Whatley, M.~Buccoli, M.~Zanoni, A.~Sarti, Virtual
  reality and choreographic practice: The potential for new creative methods,
  Body, Space \& Technology 18~(1) (2019) 1--32.

\bibitem{Kailas2021Springer}
G.~Kailas, N.~Tiwari, Design for immersive experience: Role of spatial audio in
  extended reality applications, in: A.~Chakrabarti, R.~Poovaiah, P.~Bokil,
  V.~Kant (Eds.), Design for Tomorrow---Volume 2, Springer Singapore, 2021, pp.
  853--863.

\bibitem{Majumder_2021_ICCV}
S.~Majumder, Z.~Al-Halah, K.~Grauman, Move2hear: Active audio-visual source
  separation, in: Proceedings of the IEEE/CVF International Conference on
  Computer Vision (ICCV), 2021, pp. 275--285.

\bibitem{Sundareswaran2003AR3D}
V.~Sundareswaran, K.~Wang, S.~Chen, R.~Behringer, J.~McGee, C.~Tam, P.~Zahorik,
  3{D} audio augmented reality: implementation and experiments, in: {IEEE/ACM}
  Int. Symp. on Mixed and Augmented Reality, 2003, pp. 296--297.

\bibitem{Saladnet2021}
P.~A. Grumiaux, S.~Kitić, P.~Srivastava, L.~Girin, A.~Guérin, Saladnet:
  Self-attentive multisource localization in the ambisonics domain, in: IEEE
  Workshop on Applications of Signal Process. to Audio and Acoustics (WASPAA),
  2021, pp. 336--340.

\bibitem{Mroz20216dof}
B.~Mróz, M.~Kabaciński, T.~Ciotucha, A.~Rumiński, T.~Żernicki, Production
  of six-degrees-of-freedom (6{DoF}) navigable audio using 30 ambisonic
  microphones, in: Immersive and 3D Audio: from Architecture to Automotive
  (I3DA), 2021, pp. 1--5.

\bibitem{Gotz2021Amb}
G.~Götz, S.~J. Schlecht, V.~Pulkki, A dataset of higher-order ambisonic room
  impulse responses and 3{D} models measured in a room with varying furniture,
  in: Immersive and 3D Audio: from Architecture to Automotive (I3DA), 2021, pp.
  1--8.

\bibitem{guizzo2021l3das21}
E.~Guizzo, R.~F. Gramaccioni, S.~Jamili, C.~Marinoni, E.~Massaro, C.~Medaglia,
  G.~Nachira, L.~Nucciarelli, L.~Paglialunga, M.~Pennese, S.~Pepe, E.~Rocchi,
  A.~Uncini, D.~Comminiello, {L3DAS}21 {C}hallenge: Machine learning for 3{D}
  audio signal processing, IEEE Int. Workshop on Machine Learning for Signal
  Process. (MLSP).

\bibitem{Guizzo2022L3DAS22CL}
E.~Guizzo, C.~Marinoni, M.~Pennese, X.~Ren, X.~Zheng, C.~Zhang, B.~S. Masiero,
  A.~Uncini, D.~Comminiello, {L3DAS22} challenge: Learning 3{D} audio sources
  in a real office environment, arXiv preprint: arXiv:2202.10372.

\bibitem{CAO2022141}
W.~Cao, S.~Li, J.~Zhong, {QMEDN}et: A quaternion-based multi-order differential
  encoder–decoder model for 3{D} human motion prediction, Neural Networks 154
  (2022) 141--151.

\bibitem{Zhou2022TETCI}
Y.~Zhou, L.~Jin, G.~Ma, X.~Xu, Quaternion capsule neural network with region
  attention for facial expression recognition in color images, IEEE Trans. on
  Emerging Topics in Computational Intelligence 6~(4) (2022) 893--912.

\bibitem{Huang2021QRestore}
C.~Huang, M.~K. Ng, T.~Wu, T.~Zeng, Quaternion-based dictionary learning and
  saturation-value total variation regularization for color image restoration,
  IEEE Trans. on Multimedia (2021) 1--1.

\bibitem{Jia2022TIP}
Z.~Jia, Q.~Jin, M.~K. Ng, X.-L. Zhao, Non-local robust quaternion matrix
  completion for large-scale color image and video inpainting, IEEE Trans. on
  Image Process. 31 (2022) 3868--3883.

\bibitem{SELD}
S.~Adavanne, A.~Politis, J.~Nikunen, T.~Virtanen, Sound event localization and
  detection of overlapping sources using convolutional recurrent neural
  networks, IEEE Journal of Selected Topics in Signal Process. PP (2018) 1--1.

\bibitem{ComminielloICASSP2019b}
D.~Comminiello, M.~Scarpiniti, R.~Parisi, A.~Uncini, Frequency-domain adaptive
  filtering: From real to hypercomplex signal processing, in: IEEE Int. Conf.
  on Acoust., Speech and Signal Process. (ICASSP), Brighton, UK, 2019, pp.
  7745--7749.

\bibitem{Brignone2022ISCAS}
C.~Brignone, G.~Mancini, E.~Grassucci, A.~Uncini, D.~Comminiello, Efficient
  sound event localization and detection in the quaternion domain, in: IEEE
  Trans. on Circuits and Systems II: Express Briefs, Vol.~69, 2022, pp.
  2453--2457.

\bibitem{QSSL}
M.~Ricciardi~Celsi, S.~Scardapane, D.~Comminiello, Quaternion neural networks
  for 3{D} sound source localization in reverberant environments, in: IEEE Int.
  Workshop on Machine Learning for Signal Process. (MLSP), 2020, pp. 1--6.

\bibitem{SalvatiIJCNN2020}
D.~Salvati, C.~Drioli, G.~L. Foresti, Two-microphone end-to-end speaker joint
  identification and localization via convolutional neural networks, in: 2020
  International Joint Conference on Neural Networks (IJCNN), 2020, pp. 1--6.

\bibitem{math10071083}
J.~Navarro-Moreno, R.~M. Fernández-Alcalá, J.~Ruiz-Molina, Proper {ARMA}
  modeling and forecasting in the generalized segre's quaternions domain,
  Mathematics 10~(7).

\bibitem{NavarroMorenoR21}
J.~Navarro-Moreno, J.~Ruiz-Molina.

\bibitem{Chen2021Qfact}
T.~Chen, H.~Yin, X.~Zhang, Z.~Huang, Y.~Wang, M.~Wang, Quaternion factorization
  machines: A lightweight solution to intricate feature interaction modeling,
  IEEE Trans. on Neural Netw. and Learning Systems (2021) 1--14.

\bibitem{Guo2021QHyper}
Z.~Guo, J.~Zhao, L.~Jiao, X.~Liu, F.~Liu, A universal quaternion hypergraph
  network for multimodal video question answering, IEEE Trans. on Multimedia
  (2021) 1--1.

\bibitem{Zhang2022TNNLS}
Z.~Zhang, X.~Wei, S.~Wang, C.~Lin, J.~Chen, Fixed-time pinning common
  synchronization and adaptive synchronization for delayed quaternion-valued
  neural networks, IEEE Trans. on Neural Netw. and Learning Systems (2022)
  1--14.

\bibitem{Poschadel2021EUSIPCO}
N.~Poschadel, R.~Hupke, S.~Preihs, J.~Peissig, Direction of arrival estimation
  of noisy speech using convolutional recurrent neural networks with
  higher-order ambisonics signals, in: European Signal Processing Conference
  (EUSIPCO), 2021, pp. 211--215.

\bibitem{Poppelbaum2021DQRigid}
J.~Poppelbaum, A.~Schwung, Predicting rigid body dynamics using dual quaternion
  recurrent neural networks with quaternion attention, ArXiv preprint:
  arXiv:2011.08734v1.

\bibitem{Schwung2021RigidBody}
A.~Schwung, J.~Pöppelbaum, P.~C. Nutakki, Rigid body movement prediction using
  dual quaternion recurrent neural networks, in: {IEEE} Int. Conf. on
  Industrial Technology ({ICIT}), Vol.~1, 2021, pp. 756--761.

\bibitem{Schilling2019HierarchicalDQ}
M.~Schilling, Hierarchical dual quaternion-based recurrent neural network as a
  flexible internal body model, Int. Joint Conf. on Neural Netw. (IJCNN) (2019)
  1--8.

\bibitem{TsiotrasDualQuat2020}
P.~Tsiotras, A.~Valverde, Dual quaternions as a tool for modeling, control, and
  estimation for spacecraft robotic servicing missions, J. of Astronaut. Sci.
  67 (2020) 595–629.

\bibitem{Sveier2021DualQ}
A.~Sveier, O.~Egeland, Dual quaternion particle filtering for pose estimation,
  IEEE Trans. on Control Systems Technology 29~(5) (2021) 2012--2025.

\bibitem{Gui2021DualQ}
H.~Gui, Y.~Wang, W.~Su, Hybrid global finite-time dual-quaternion observer and
  controller for velocity-free spacecraft pose tracking, IEEE Trans. on Control
  Systems Technology 29~(5) (2021) 2129--2141.

\bibitem{CaoDual2021}
Z.~Cao, Q.~Xu, Z.~Yang, X.~Cao, Q.~Huang, Dual quaternion knowledge graph
  embeddings, Proc. of the {AAAI} Conference on Artificial Intelligence 35~(8)
  (2021) 6894--6902.

\bibitem{Nguyen2022NodeCB}
D.~Q. Nguyen, V.~Van~Tong, D.~Q. Phung, D.~Q. Nguyen, Node co-occurrence based
  graph neural networks for knowledge graph link prediction, {ACM} Int. Conf.
  on Web Search and Data Mining.

\bibitem{Plinge2018SixDegreesofFreedomBA}
A.~Plinge, S.~J. Schlecht, O.~Thiergart, T.~Robotham, O.~S. Rummukainen,
  E.~A.~P. Habets, Six-degrees-of-freedom binaural audio reproduction of
  first-order ambisonics with distance information, in: {AES} Int. Conf. on
  Audio for Virtual and Augmented Reality ({AVAR}), 2018.

\bibitem{Ward1997}
J.~P. Ward, Quaternions and Caley Numbers. Algebra ans Applications, Vol. 403
  of Mathematics and Its Applications, Kluwer Academic Publishers, 1997.

\bibitem{Kuipers1921book}
J.~B. Kuipers, Quaternions and rotation sequences, Princeton University Press,
  1921.

\bibitem{Clifford1871}
M.~A. Clifford, Preliminary sketch of biquaternions, Proceedings of the London
  Mathematical Society s1-4~(1) (1871) 381--395.

\bibitem{Jia2018DualQ}
Y.-B. Jia, Dual quaternions, Iowa State University, 2018.

\bibitem{Samanc2021TheNS}
Samancı, H. K. and Kuşçu, C., The New
  Screw Interpolations and Their Geometric Properties in the Dual Spherical
  Mechanisms, Uluslararası Muhendislik Arastirma ve
  Gelistirme Dergisi.

\bibitem{Valverde2018DynamicMA}
Valverde, A., Dynamic modeling and control
  of spacecraft robotic systems using dual quaternions, in:
  PhD Thesis, Georgia Institute of
  Technology, 2018.

\bibitem{McCarthyDualQ1990}
McCarthy, J. M., An Introduction to
  Theoretical Kinematics, The MIT Press,
  1990.

\bibitem{GrassucciICASSP2021}
E.~Grassucci, D.~Comminiello, A.~Uncini, A quaternion-valued variational
  autoencoder, in: IEEE Int. Conf. on Acoust., Speech and Signal Process.
  ({ICASSP}), Toronto, Canada, 2021.

\bibitem{GrassucciQGAN2021}
E.~Grassucci, E.~Cicero, D.~Comminiello, Quaternion generative adversarial
  networks, in: R.~Razavi-Far, A.~Ruiz-Garcia, V.~Palade, J.~Schmidhuber
  (Eds.), Generative Adversarial Learning: Architectures and Applications,
  Springer International Publishing, Cham, 2022, pp. 57--86.

\bibitem{SELD-TCN}
K.~Guirguis, C.~Schorn, A.~Guntoro, S.~Abdulatif, B.~Yang, {SELD-TCN}: Sound
  event localization \& detection via temporal convolutional networks, Europ.
  Signal Process. Conf. ({EUSIPCO}).

\bibitem{oord2016wavenet}
A.~Van Den~Oord, S.~Dieleman, H.~Zen, K.~Simonyan, O.~Vinyals, A.~Graves,
  N.~Kalchbrenner, A.~Senior, K.~Kavukcuoglu, Wavenet: A generative model for
  raw audio, ArXiv preprint: arXiv:1609.03499.

\bibitem{wavenet}
D.~Rethage, J.~Pons, X.~Serra, A wavenet for speech denoising, 2018, pp.
  5069--5073.

\bibitem{grassucci2021lightweight}
E.~Grassucci, A.~Zhang, D.~Comminiello, Phnns: Lightweight neural networks via
  parameterized hypercomplex convolutions, ArXiv preprint: arXiv:2110.04176.

\bibitem{Fonseca2022FSD50KAO}
E.~Fonseca, X.~Favory, J.~Pons, F.~Font, X.~Serra, Fsd50k: An open dataset of
  human-labeled sound events, {IEEE/ACM} Trans. on Audio, Speech, and Language
  Process. 30 (2022) 829--852.

\bibitem{Politis2021OverviewAE}
A.~Politis, A.~Mesaros, S.~Adavanne, T.~Heittola, T.~Virtanen, Overview and
  evaluation of sound event localization and detection in dcase 2019,
  {IEEE/ACM} Trans. on Audio, Speech, and Language Process. 29 (2021) 684--698.

\bibitem{Mesaros2019Measurement}
A.~Mesaros, S.~Adavanne, A.~Politis, T.~Heittola, T.~Virtanen, Joint
  measurement of localization and detection of sound events, in: {IEEE}
  Workshop on Applications of Signal Process. to Audio and Acoustics
  ({WASPAA}), 2019, pp. 333--337.

\end{thebibliography}

\end{document}